\documentclass[apj]{emulateapj}
\usepackage{apjfonts}
\usepackage{graphicx}

\begin{document}

\title{SIMULATION STUDY OF DUST-SCATTERED FAR-ULTRAVIOLET EMISSION IN THE ORION-ERIDANUS SUPERBUBBLE}

\author{Young-Soo Jo\altaffilmark{1}, Kyoung-Wook Min\altaffilmark{1}
, Tae-Ho Lim\altaffilmark{1}, Kwang-Il Seon\altaffilmark{2}}

\email{email: stspeak@kaist.ac.kr}

\altaffiltext{1}{Korea Advanced Institute of Science and Technology
(KAIST), 373-1 Guseong-dong, Yuseong-gu, Daejeon, Korea 305-701,
Republic of Korea}

\altaffiltext{2}{Korea Astronomy and Space Science Institute (KASI),
61-1 Hwaam-dong, Yuseong-gu, Daejeon, Korea 305-348, Republic of
Korea}

\begin{abstract}
We present the results of dust scattering simulations carried out
for the Orion Eridanus Superbubble region by comparing them with
observations made in the far-ultraviolet. The albedo and the phase
function asymmetry factor (\textit{g}-factor) of interstellar grains
were estimated, as were the distance and thickness of the dust
layers. The results are as follows: 0.43$^{+0.02}_{-0.04}$ for the
albedo and 0.45$^{+0.2}_{-0.2}$ for the \textit{g}-factor, in good
agreement with previous determinations and theoretical predictions.
The distance of the assumed single dust layer, modeled for the Orion
Molecular Cloud Complex, was estimated to be $\sim$110 pc and the
thickness ranged from $\sim$130 at the core to $\sim$50 pc at the
boundary for the region of the present interest, implying that the
dust cloud is located in front of the superbubble. The simulation
result also indicates that a thin ($\sim$10 pc) dust shell surrounds
the inner X-ray cavities of hot gas at a distance of $\sim$70-90 pc.
\end{abstract}


\section{INTRODUCTION}

As stellar ultraviolet (UV) radiations scattered by interstellar
dust grains are regarded as the most dominant source of the diffuse
Galactic light \citep{seo11}, the optical properties of the dust
grains, generally characterized by the albedo and the phase function
asymmetry factor \textit{g} $\equiv$ $<\cos\theta>$, can be inferred
from the observations of the diffuse Galactic light. The
observations can also be compared with model calculations. For
example, \citet{dra03} estimated the albedo and \textit{g}-factor
for the carbonaceous-silicate grains, and obtained $\sim$0.40 and
$\sim$0.65 for the albedo and the \textit{g}-factor at $\lambda$
$\sim$ 1600 {\AA}, respectively. While the values of the albedo and
g-factor are clearly dependent on the models of dust grains,
accurate photometry and modeling of dust clouds are also required
for reliable estimations of these optical properties from
observations. Until now, there have been numerous studies in this
line of research from the early 70's, especially at far-ultraviolet
(FUV) and near-ultraviolet (NUV) wavelengths
\citep{wit73,wit78,lil76,mor78,mor82,mor80,ona84,wit92,wit97,gor94,cal95}.
For example, the observations made from the \textit{Voyager 2}
spacecraft have been used for the wavelengths shorter than 1200
{\AA} \citep{wit93,mur93,bur02,sha04,sha06,suj05,suj07}. A
comparative review of past work can be found in \citet{gor04}.
Recently, \textit{GALEX} observations provided diffuse UV background
images with high spatial resolution, which have been used for the
determination of the albedo and the \textit{g}-factor in both the
FUV (1350-1750 {\AA}) and NUV (1750-2850 {\AA}) bands
\citep{suj09,suj10}. Most of these studies showed that the
interstellar grain causes a strong forward scattering in the FUV,
with a moderate albedo, which is in good agreement with theoretical
predictions.

The Orion-Eridanus Superbubble (henceforth OES) is a large hot
bubble with a diameter of $\sim$30$^\circ$ across the sky located in
the Orion and Eridanus constellations. Its distance is known to be
$\sim$155 pc to the near-side and $\sim$586 pc to the far-side of
the bubble, and its near edge is considered to interact with the
expanding shell of the Local Bubble \citep{bur96}. The OES has been
observed extensively in various wavelengths, ranging from X-ray to
radio. Enhanced soft X-ray emission seen at its inner cavities is
the evidence of the hot gas (T $\sim$ 10$^6$ K) occupying its inner
parts \citep{hei99}. Two prominent filamentary H$\alpha$ shells,
known as arcs A and B, have been observed at the boundaries of these
hot cavities \citep{rey98,bou01}. The presence of 21 cm radiation of
neutral hydrogen (T $\sim$ 10$^{2-3}$ K) was also noted outside the
hot bubble (Brown et al. 1995). The dust extinction
(\textit{E}(\textit{B-V})) map, made from the Schlegel, Finkbeiner
and Davis (henceforth SFD) Dust Survey \citep{sch98}, shows a strong
anti-correlation with the soft X-ray emission image \citep{jo11}.
Recent spectroscopic observations made in the FUV revealed the
existence of \ion{C}{4} and \ion{Si}{2}* emission lines as well as
molecular fluorescence H$_2$ lines, implying that the OES is truly a
multi-phase object, consisting of hot gases and cold dust that
interact with each other \citep{kre06,ryu06,ryu08,jo11}.

The dust properties of the OES region have not been explored well
because of the lack of sufficient diffuse FUV observations. Based on
the data available from the Voyager measurement, \citet{mur93}
suggested a lower limit of $\sim$0.3 for the albedo assuming
isotropic scattering, and an upper limit of $\sim$0.8 for the phase
function asymmetry factor with perfectly reflecting grains. However,
it should be noted that the Voyager observations provided limited
information, and only for two regions with a field of view of
0$^\circ$.10 $\times$ 0$^\circ$.87 and a spectral band of 38 {\AA}.
In this paper, we report the results of our study on the dust
scattering properties of the OES region. We performed Monte Carlo
simulations and compared the results with the diffuse emission map
made from the recent FUV imaging spectrograph mission. We obtained
the average albedo and phase function asymmetry factor, and
estimated the distance and thickness of the dust cloud.

\begin{figure}
 \begin{center}
  \includegraphics[width=7.3cm]{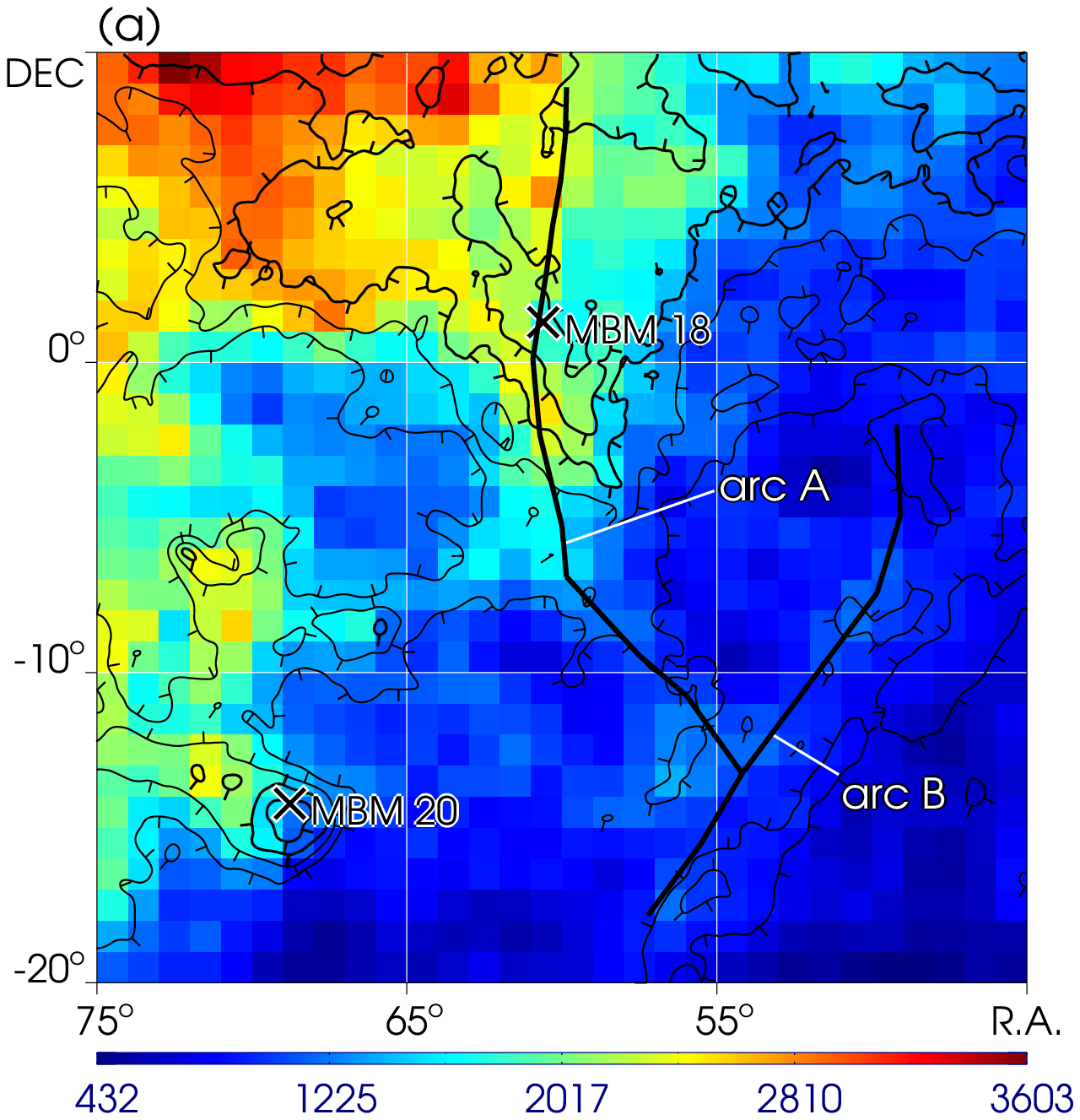}\\
  \includegraphics[width=7cm]{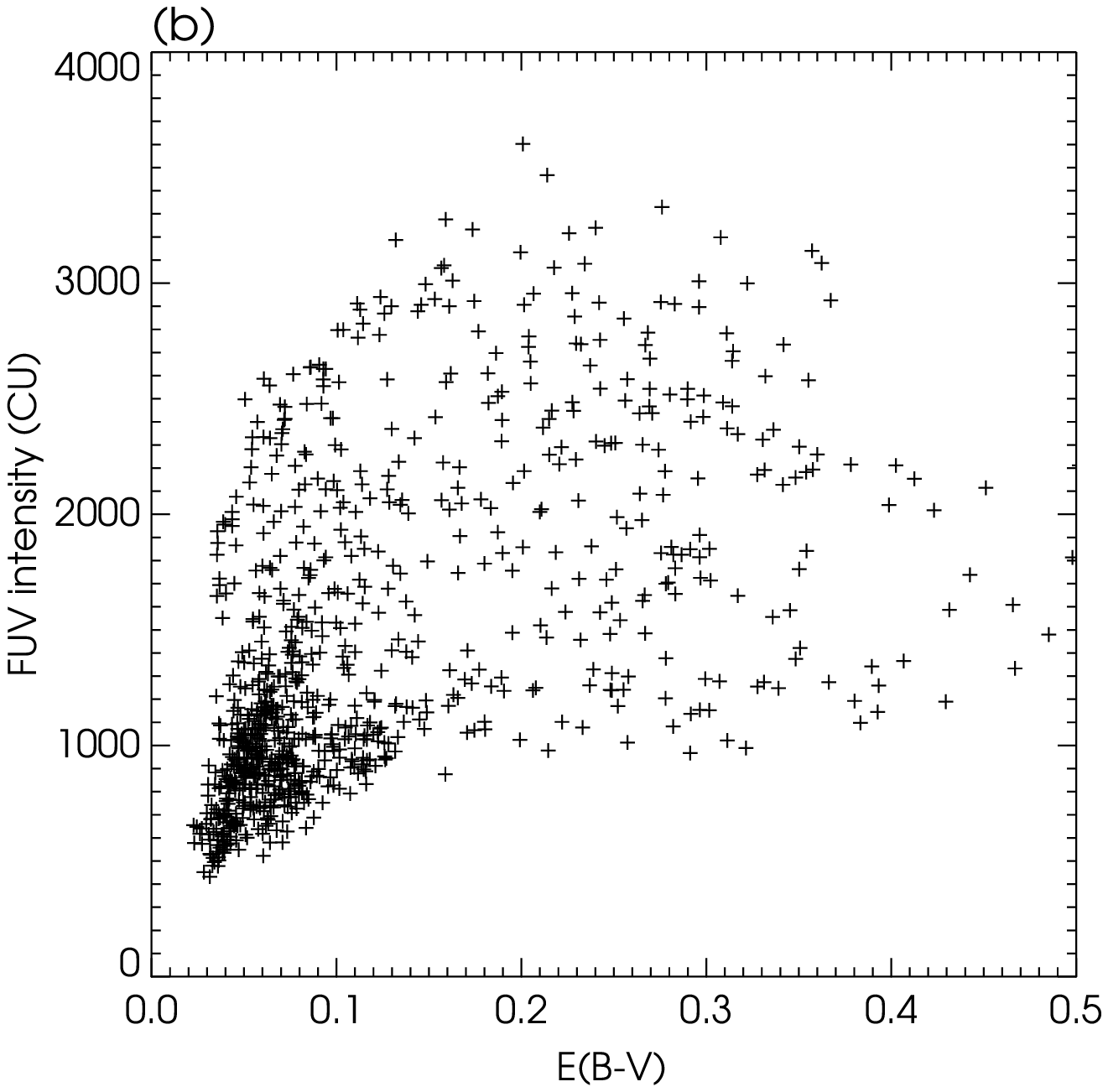}\\
 \end{center}
\caption{(a) FUV continuum map overplotted with dust contours, in
equatorial coordinates. The continuum intensity is given in CU
(photons s$^{-1}$ cm$^{-2}$ sr$^{-1}$ {\AA}$^{-1}$) and the levels
of dust extinction contours are in \textit{E}(\textit{B-V}). (b) A
scatter plot for the FUV intensity against dust extinction, obtained
from pixel-by-pixel comparison of Figure \ref{fig:fims}(a).
\label{fig:fims}}
\end{figure}


\section{Observations and Data Reduction}

The present study is based on the same data set obtained via
Far-Ultraviolet Imaging Spectrograph (FIMS) as was used previously
in the studies of \citet{ryu06,ryu08} and \citet{jo11}. FIMS is a
dual-channel FUV imaging spectrograph (S-band 900 {\AA}-1150 {\AA},
L-band 1350 {\AA}-1750 {\AA}) on board the Korean microsatellite
STSAT-1, which was launched into a 700 km sun-synchronous orbit on
September 27, 2003 and operated for a year and a half. With its
large image fields of view (S-band: 4$^\circ$.0$\times$4'.6, L-band:
7$^\circ$.5$\times$4'.3) and moderate spectral
($\lambda/\Delta\lambda$ $\sim$ 550) and angular ($\sim$5')
resolutions, the instrument is optimized for the observation of
diffuse emissions \citep{ede06a,ede06b}. Though the OES region was
observed in both the S- and L-band, we analyzed only the L-band data
for the present study since the S-band data were strongly
contaminated by geocoronal emission lines. Bright point sources were
removed from the map: a total of 210 stars in the TD-1 catalog with
1565 {\AA} band fluxes  of above 10$^{-12}$ ergs {\AA}$^{-1}$
s$^{-1}$ cm$^{-2}$ were removed, and the remaining bright pixels,
not corresponding to the stars listed in the TD-1 catalog, were also
removed. More information on the data reduction procedures can be
found in \citet{jo11}.

While stellar UV radiation scattered by interstellar dust grains is
generally known to be the most dominant source of diffuse FUV
emission, there are other contributors such as atomic line
emissions, extragalactic background, H$_2$ fluorescence, and
two-photon continuum emissions. Hence, it is necessary to eliminate
these extra contributions in order to obtain more reliable results
for the effect of dust scattering. For example, the atomic lines
were removed as follows. We took only the 1420 {\AA}-1630 {\AA}
portion from the L-band spectrum to exclude the bright airglow lines
at \ion{O}{1} 1356 {\AA} and 1641 {\AA} as well as other
atomic/ionic lines such as \ion{O}{4}] 1407 {\AA}, \ion{He}{2} 1640
{\AA}, \ion{C}{1} 1657 {\AA}, and \ion{Al}{2} 1671 {\AA}, as
identified previously in Table 1 of \citet{jo11}. The remaining
spectrum still included the \ion{Si}{2}* and \ion{C}{4} lines
centered at 1533 {\AA} and 1549 {\AA}, respectively, which were
removed by excluding the spectral range from 1525 {\AA} to 1560
{\AA}. We have constructed a diffuse FUV map with the remaining
portion of the spectrum.

The extragalactic diffuse FUV background is known to be more or less
uniform over the celestial sphere with $\sim$300 Continuum Units
(CU), and is expected to be even more so at high galactic latitudes
\citep{bow91}. Noting that OES is located at high galactic
latitudes, we have subtracted 300 CU from the original diffuse
emission map. The contribution of H$_2$ fluorescent emission was
subtracted in the following way using the map of Figure 2(b) in
\citet{jo11}. We first converted the original H$_2$ map, which was
constructed from the 1608 {\AA} line intensity, to a map of total
H$_2$ emission for the whole wavelength range in consideration of
using the H$_2$ fluorescence emission profiles given by the CLOUD
code \citep{van86,bla87}, a simulation model for a photodissociation
region (PDR). We assumed that the intensity ratios among the lines
were the same for the entire OES region. The resulting emission
intensity was expressed in terms of CU by dividing the total
intensity with the wavelength range, and subtracted from the diffuse
emission map.

It was mentioned in \citet{jo11} that two-photon emission
constitutes about 50$\%$-70$\%$ of the FUV continuum intensity in
the lower part of arc A and arc B, based on the assumption that 1
Rayleigh of the H$\alpha$ line corresponds to the two-photon (plus
free-free and bound-free) continuum emission of 60 CU in an ionized
gas \citep{rey90}. On the other hand, \citet{seo11} recently
suggested that the contribution of two-photon effects to the FUV
continuum could be lower in the case of cooling ionized gas with 1
Rayleigh of the H$\alpha$ line corresponding to $\sim$26.0 CU.
Furthermore, Figure 6 of \citet{wit10}, which shows the data for the
OES region studied by \citet{mad06}, suggests that 10 to 40$\%$ of
the total H$\alpha$ intensity comes from dust scattering. These
complexities make it difficult to estimate the effect of two-photon
emissions.  Therefore, we have not removed the two-photon continuum
emission, which may make the two arcs appear more prominently, and
will instead discuss its contribution to the map. The final image
was obtained with 1$^\circ$ $\times$ 1$^\circ$ pixel bins and is
shown in Figure \ref{fig:fims}(a), overplotted with dust contours.

When the new map of Figure \ref{fig:fims}(a) is compared with the
previous diffuse FUV map, Figure 2(a) of \citet{jo11}, which
includes the contributions of the ion lines as well as the H$_2$
fluorescent emission, we can see that overall morphology did not
change much. The most significant change due to the exclusion of the
ion lines is seen only in the northeastern region of the map near
the left boundary above $\delta$ = 0$^\circ$, where the scattered
photons of the two bright B-type stars with strong P Cygni profiles,
HD 31237 and HD 30836 located at ($\alpha$, $\delta$) $\sim$
(73$^\circ$.6, 2$^\circ$.4) and (72$^\circ$.8, 5$^\circ$.6),
respectively, are responsible for the enhanced \ion{C}{4} emissions.
The H$_2$ fluorescent emission varied significantly according to the
regions as can be seen in Figure 2(b) of \citet{jo11} while it is
$\sim$17$\%$ on average for the whole OES region, as can be
estimated from the spectrum of Figure 1 in \citet{jo11}. The
contribution of H$_2$ fluorescent emission amounts to $\sim$35$\%$
of the continuum intensity near the western edge of the lower part
of arc A. The intensity levels of this region decreased from
$\sim$2,000 CU to $\sim$1,200 CU after subtraction of H$_2$
emission. Two-photon effects of hydrogen atom are clearly seen in
the two strong H$\alpha$ filamentary structures, arc A and arc B,
whose centroids are depicted as thick black solid lines in Figure
\ref{fig:fims}(a).

As Figure \ref{fig:fims}(a) generally shows high FUV intensity in
the regions where the color excess \textit{E}(\textit{B-V}) is
large, we made a pixel-to-pixel scatter plot of the FUV intensity
against dust extinction in Figure \ref{fig:fims}(b) to check the
correlation between them. A correlation is seen for
\textit{E}(\textit{B-V}) < 0.14, especially for the FUV intensity
less than 1500 CU, while FUV intensity becomes scattered and more or
less independent of dust extinction for \textit{E}(\textit{B-V})
above 0.14. In fact, it should be noted that the color excess value
of $\sim$0.14 corresponds to the optical depth of $\tau$ $\sim$ 1 at
1565 {\AA}. The result agrees well with the general notion that the
FUV continuum and dust extinction have a correlation in an optically
thin region. The data points of \textit{E}(\textit{B-V}) < 0.14 with
its FUV intensity of less than 1500 CU correspond to the blue region
in Figure \ref{fig:fims}(a): among these data points, those of
higher FUV intensity correspond to the inner region of the OES shell
while those of lower FUV intensities correspond to the outer regions
of the OES shell. Both of them seem to show a good linear
relationship, implying more or less constant radiation fields.

The data points of \textit{E}(\textit{B-V}) < 0.14 with its FUV
intensity of above 1500 CU generally correspond to the regions of
light blue to yellow around the boundary of the thick dust region.
These regions show high FUV intensity due to the strong UV radiation
fields produced by the nearby B type stars. The strong UV radiations
from these bright B type stars also affect the optically thick
region in the eastern part of the Orion Molecular Cloud Complex
(henceforth OMCC): the data points of \textit{E}(\textit{B-V}) >
0.14 and the FUV intensity above 2500 CU correspond to the regions
of yellow to red in the eastern part of the Complex in Figure
\ref{fig:fims}(a). The data points of \textit{E}(\textit{B-V})
> 0.14 and FUV intensity < 2500 CU correspond to the regions of
light blue to green in the western part of the OMCC. As noted in
\citet{jo11}, the western part of the Complex shows lower FUV
intensity than its eastern counterpart because it contains fewer
stars as interstellar radiation sources. Hence, we see a linear
relationship between FUV intensity and dust extinction in the blue
regions of Figure \ref{fig:fims}(a) where dust extinction is small
while the scattered data points in Figure \ref{fig:fims}(b) seem to
be related to the fluctuations in the background UV radiation
fields.

\begin{figure}
 \begin{center}
  \includegraphics[width=7.3cm]{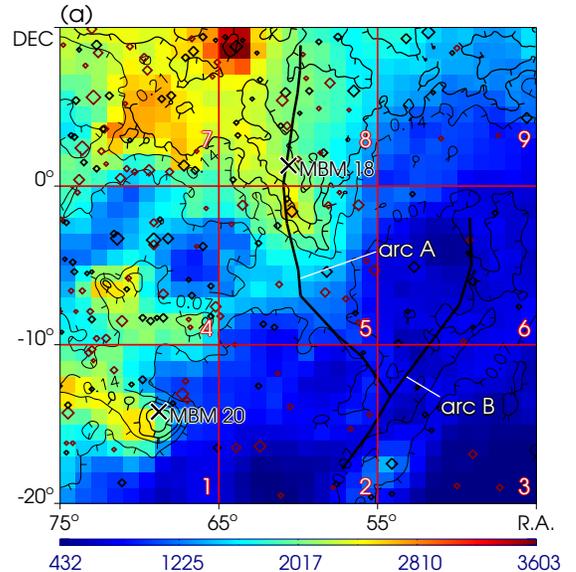}\\
  \includegraphics[width=7cm]{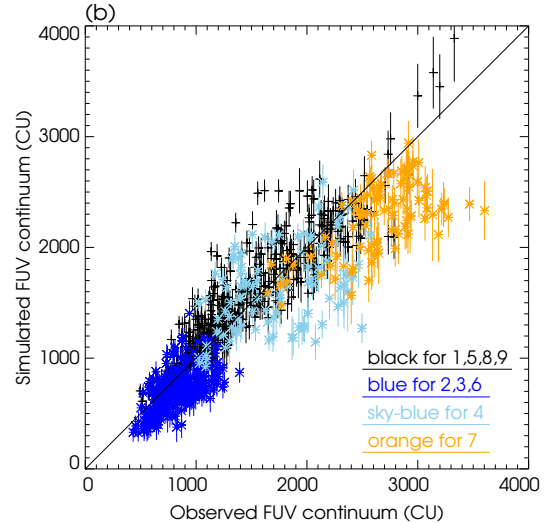}\\
 \end{center}
\caption{(a) FUV continuum map obtained from dust scattering
simulation for a single dust slab, and (b) a pixel-to-pixel scatter
plot of the FUV intensities for the observed and simulated maps in
Figure \ref{fig:fims}(a) and Figure \ref{fig:simul}(a),
respectively. 194 stars selected as photon sources are marked in
(a): black symbols correspond to the stars closer than 200 pc, and
red symbols correspond to the stars located farther than 200 pc. The
symbol size is proportional to the 1565 {\AA} band flux of the TD-1
star catalog in log scale. \label{fig:simul}}
\end{figure}

\begin{figure*}
 \begin{center}
  \includegraphics[width=5.2cm]{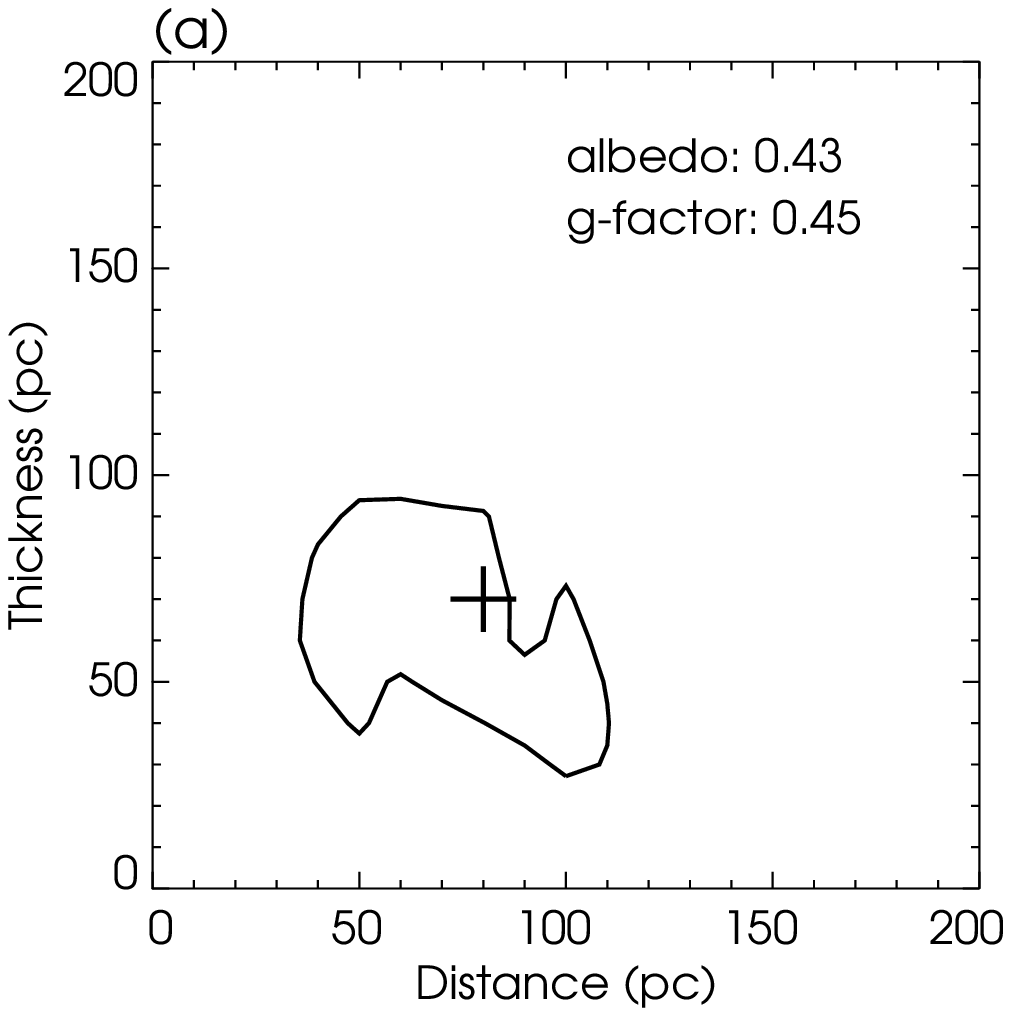}\hspace{0.8cm}
  \includegraphics[width=5.2cm]{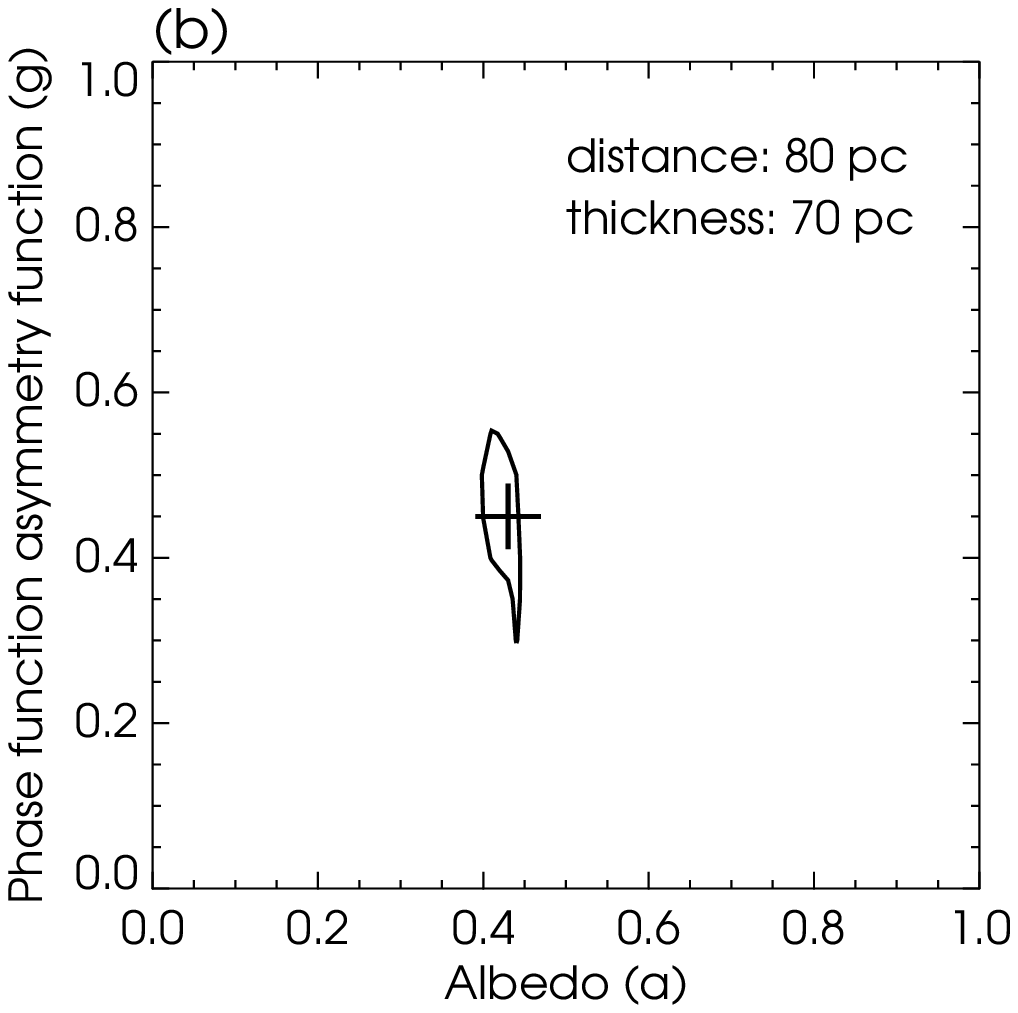}\hspace{0.8cm}
  \includegraphics[width=5.2cm]{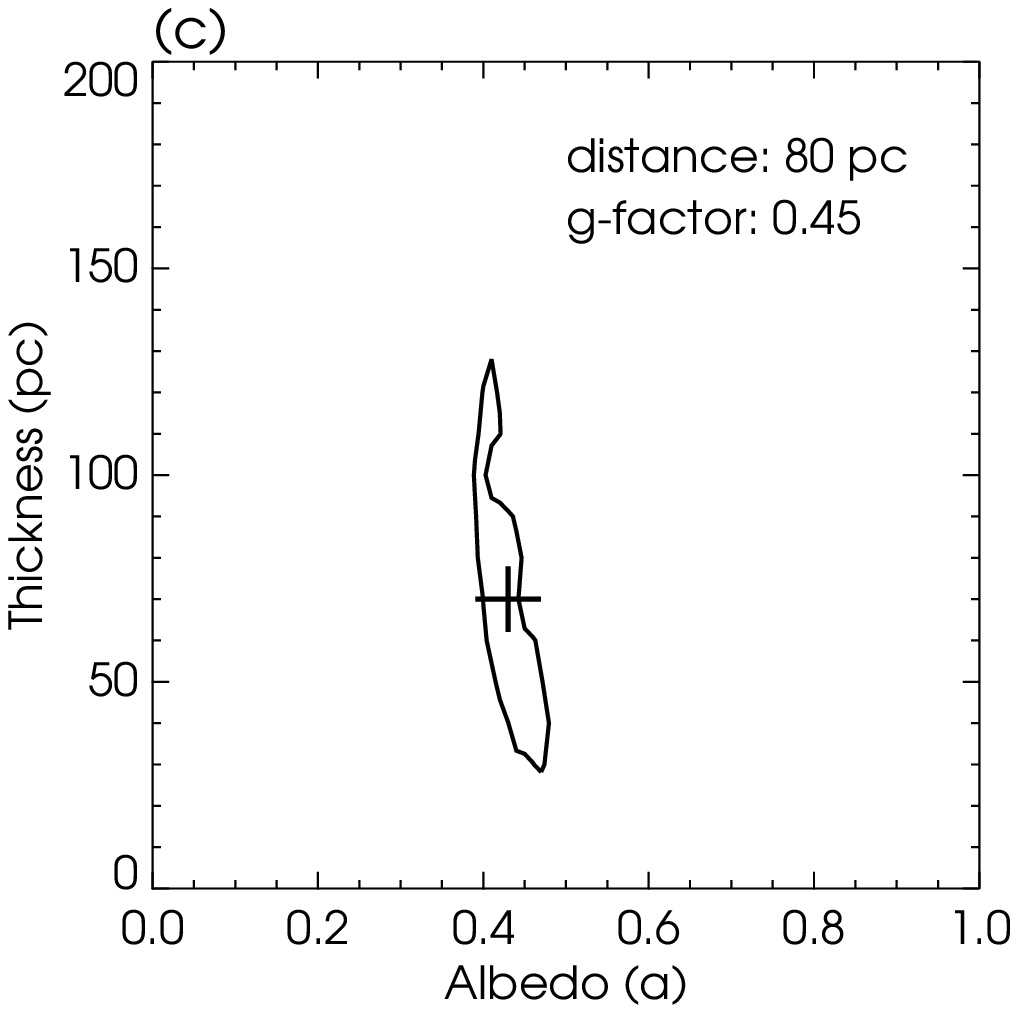}\\
  \vspace{0.8cm}
  \includegraphics[width=5.2cm]{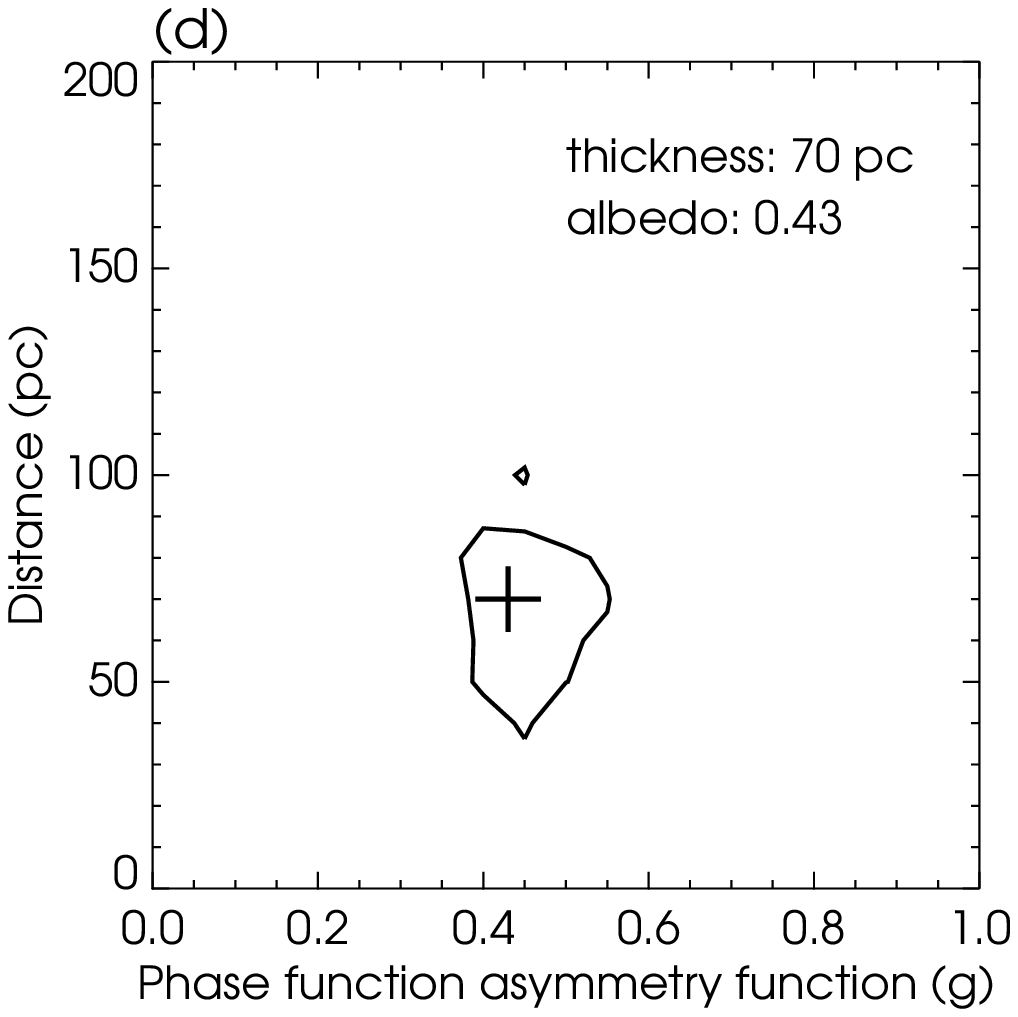}\hspace{0.8cm}
  \includegraphics[width=5.2cm]{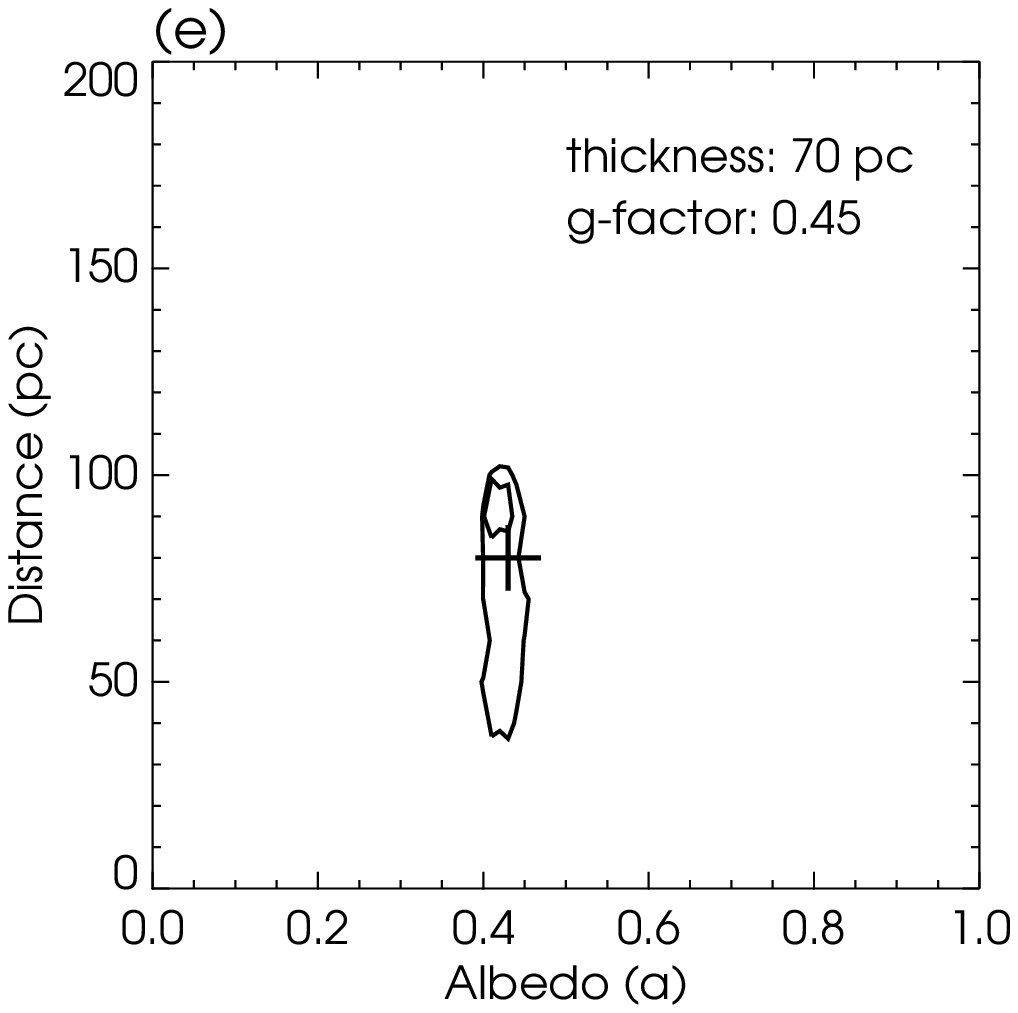}\hspace{0.8cm}
  \includegraphics[width=5.2cm]{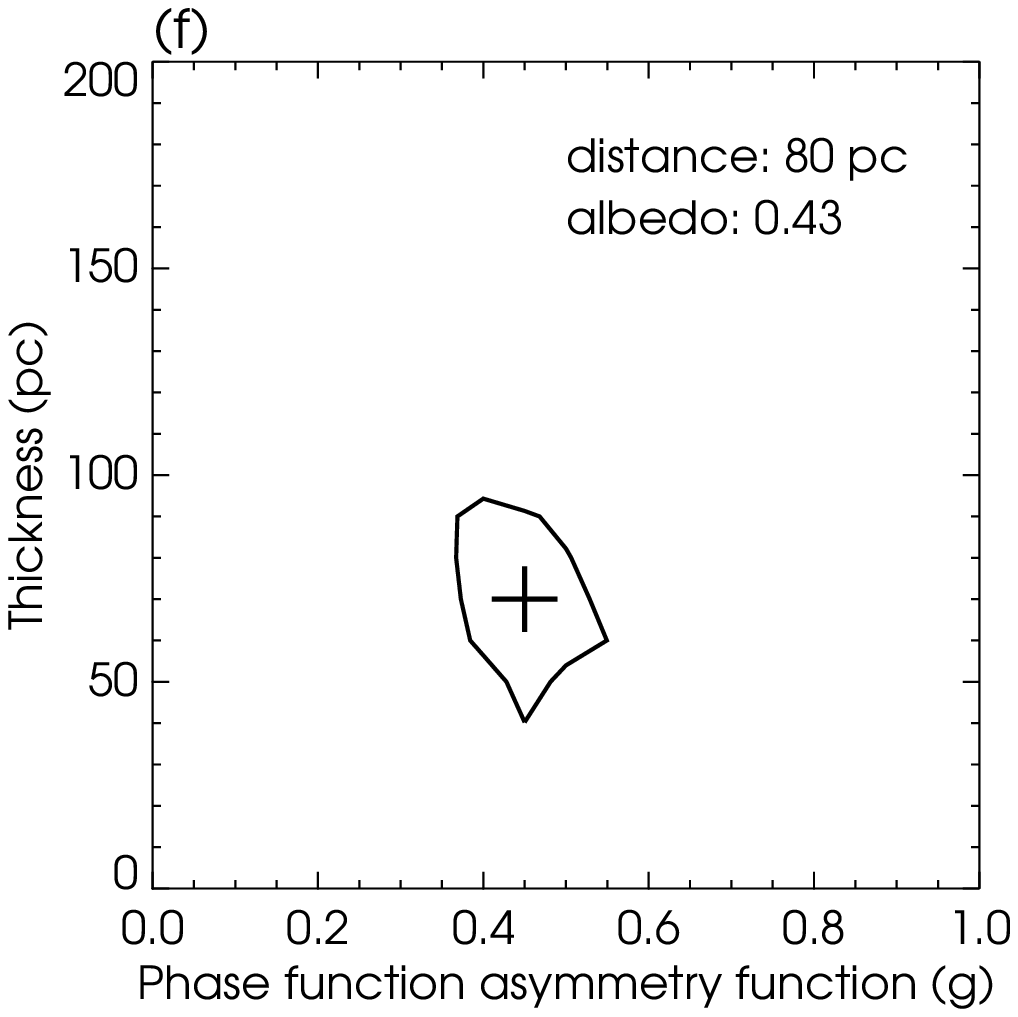}\\
 \end{center}
\caption{90$\%$ confidence contours: (a) thickness vs. distance
plotted for the albedo 0.43 and the \textit{g}-factor 0.45, (b)
g-factor vs. albedo plotted for the distance 80 pc and the thickness
70 pc, (c) thickness vs. albedo plotted for the distance 80 pc and
the \textit{g}-factor 0.45, (d) distance vs. \textit{g}-factor
plotted for the thickness 70 pc and the albedo 0.43, (e) distance
vs. albedo plotted for the thickness 70 pc and the \textit{g}-factor
0.45, and (f) thickness vs. \textit{g}-factor plotted for the
distance 80 pc and the albedo 0.43. Here, distance indicates the
distance to the front face of the slab. \label{fig:contour}}
\end{figure*}


\section{Simulation Model and Result}

As the observed FUV intensity generally follows the dust extinction
level in Figure \ref{fig:fims}(a), with fluctuations which may be
attributable to the local background UV radiation fields, most of
the FUV diffuse emission in the OES region seems to originate from
dust scattering of the radiation fields produced by nearby stars.
Hence, it might be possible to obtain useful information regarding
the distribution and scattering properties of dust from the observed
FUV continuum map by comparing it with simulations based on
theoretical predictions. We have performed Monte Carlo simulations,
in which photons originating from bright stars are multiply
scattered by assumed distributions of dust. We adopted the following
Henyey-Greenstein scattering phase function $\Phi$($\theta$)
\citep{hen41} for the dust scattering model, which is relatively
simple and known to fit the observations reasonably well.
\begin{equation}
    \phi(\theta) = \frac{a}{4\pi}\frac{(1-g^2)}{[1+g^2-2g\cos(\theta)]^{1.5}}
\end{equation}
where \textit{a} and \textit{g} are the albedo and phase function
asymmetry factor of dust grains, respectively, and $\theta$ is the
scattering angle measured from the direction of the incident photon.
The simulation code adopted the peeling-off method \citep{yus84}:
while each photon experiences multiple scatterings in random
directions as it propagates through the dust cloud, the code keeps
track of the fraction of a photon in the direction toward the Sun
from each scattering, which is summed up for the final intensity.
The Monte Carlo simulation code is described in more detail in
\citet{seo09} and Seon et al. (2012, to be submitted).

We took a 400 pc $\times$ 400 pc $\times$ 800 pc rectangular box as
a simulation domain, with 800 pc chosen to be the distance to the
rear face from the Sun in view of the size of the OES region whose
distance is known to be $\sim$155 pc at the near-side and $\sim$586
pc at the far-side. The central axis was chosen along the line of
sight toward the center of OES, ($\alpha$, $\delta$) = (60$^\circ$,
-5$^\circ$), with the Sun placed at the center of the front face.
The 400 pc $\times$ 400 pc square located at 750 pc from the Sun
corresponds to 30$^\circ$ $\times$ 30$^\circ$ angular span in the
sky, encompassing the entire OES region of 30$^\circ$ $\times$
30$^\circ$ of the present study.  The bright stars of the TD-1 star
catalog, whose 1565 {\AA} band flux is greater than 10$^{-12}$ ergs
{\AA}$^{-1}$ s$^{-1}$ cm$^{-2}$, were selected as the point sources
of stellar photons. The total number of stars is 3,338 for the whole
simulation box, 194 of which belong to the region of a 30$^\circ$
$\times$ 30$^\circ$ angular span of the present interest and are
shown in Figure \ref{fig:simul}(a). Incorporation of the additional
\textit{Hipparcos} stars did not change the result much, though the
\textit{Hipparcos} catalog contains $\sim$3,000 more stars than the
TD-1 catalog for the OES region: the total luminosity increase was
only $\sim$1$\%$. The stellar luminosities were calculated with the
distances adopted from the \textit{Hipparcos} star catalog, together
with the extinction corrections made by comparing the observed (B-V)
values and the model (B-V) values derived from the Castelli model
\citep{per97,cas03}. We checked the \textit{E}(\textit{B-V}) values
for the 19 OES stars listed in the catalog of \citet{fri92} and
found good agreement within 15$\%$ on average for sufficiently high
\textit{E}(\textit{B-V}) as much as 0.02. The optical depth at 1565
{\AA} was derived from \textit{E}(\textit{B-V}) assuming R$_V$ to be
= 3.1, an average value of the diffuse ISM, with the absorption
coefficients calculated for the carbonaceous-silicate grains model
\citep{wei01}. In fact, we estimated the average R$_V$ value for the
OES region using the values made for the 6 OES stars in
\citet{lar05} and the A$_V$ values provided by \citet{nec80} for the
11 OES stars with \textit{E}(\textit{B-V}) > 0.05, and obtained 3.2,
which was very close to the average value of the diffuse ISM.
Furthermore, the extinction curves for five OES stars (HD23466,
24263, 26912, 30076, and 30870) in \citet{pap91} are more or less
consistent with R$_V$=3.1.

The most serious difficulty with the present simulation is related
to the dust distribution since there is no report regarding the
spatial distribution of dust for the OES region, other than the
infrared survey results that give only indirect information on the
integrated amount of dust along the sightlines. Hence, we assumed
the dust cloud to be a slab of finite thickness for the sake of
convenience. This idealized dust model may certainly be different
from the real dust distribution, and the result of our simulations
should be regarded as a representative one, indicating the effective
distance and thickness. In the present simulation, the
\textit{E}(\textit{B-V}) values of the SFD dust map for given
sightlines were converted into the optical depths at 1565 {\AA},
which was regarded as a representative FUV wavelength of the FIMS
observation, and then assigned to the grid points with equal
distribution for an assumed cloud thickness. As we learned that most
of the dust is located within $\sim$400 pc from the Sun in the
direction toward the OES (\citet{ver10}; Vergely 2011 private
communication), we took a single dust slab and placed it within 400
pc with the whole dust extinction from the SFD map distributed
equally among the grids for an assumed thickness. We will further
consider possible improvements to our model in the Discussion
section.

Along with the distance and thickness of the dust slab, the albedo
and phase function asymmetry factor, which are the optical
parameters related to the scattering properties of dust, were
varied. The ranges of simulation parameters were as follows: the
distance to the front face varied from 10 pc to 200 pc with 10 pc
steps, the thickness varied from 10 pc to 200 pc with 10 pc steps,
the albedo varied from 0.35 to 0.50 with 0.01 steps, and the
g-factor varied from 0.20 to 0.70 with 0.05 steps. For each
simulation, 10$^8$ photons in total were randomly generated from the
point sources. Each simulation result was compared with the FIMS
observation to obtain the best fitting parameters using the
chi-square minimization method.

Figure \ref{fig:simul}(a) shows the simulated FUV map for the best
fitting parameters: 80 pc for the distance to the front face of the
dust slab, 70 pc for the thickness of the slab, 0.43 for the albedo,
and 0.45 for the \textit{g}-factor. Figure \ref{fig:simul}(b) is a
scatter plot obtained from a pixel-to-pixel comparison between the
FIMS map of Figure \ref{fig:fims}(a) and the simulation map of
Figure \ref{fig:simul}(a). We have divided Figure \ref{fig:simul}(a)
into nine sub-regions with a 10$^\circ$ $\times$ 10$^\circ$ area and
numbered the sub-regions from 1 to 9, so that the data points in
Figure \ref{fig:simul}(b) were color-coded according to the
sub-regions. It can be seen in Figure \ref{fig:simul}(b) that the
simulated intensity generally agrees with the observed one. The
reduced chi-square value is $\chi^2$ $\sim$ 5.44. A little lower
simulated intensity than that observed below 1500 CU of FIMS
intensity (blue) corresponds to the regions of arc B and the lower
part of arc A, where two-photon emission, not considered in the
simulation, is a significant contribution. If the difference is
regarded as the contribution from two-photon emissions, the amount
is estimated to be $\sim$25-40$\%$ of the total observed continuum,
a little lower than the previous value of $\sim$50-70$\%$ in
\citet{jo11}, which was obtained by converting the observed
H$\alpha$ intensity assuming that 1 Rayleigh of the H$\alpha$ line
corresponded to the two photon emission of 60 CU.

Large fluctuations are seen between 1500 and 2500 CU of FIMS
intensity (sky-blue), and the simulation result is considerably
lower than the observed one above 2500 CU of FIMS intensity
(orange). These discrepancies may originate from local variations of
the simulation parameters such as the distance to the front face and
the thickness of the dust cloud. More consideration of these
discrepancies will be given in the Discussion section.

Figure \ref{fig:contour} shows the contour maps for the 90 $\%$
confidence ($\Delta\chi^2$ $\sim$ 1.06) among the four variables,
the albedo, the \textit{g}-factor, the distance to the front face of
the slab, and the thickness: the ranges are 0.43$^{+0.02}_{-0.04}$
for the albedo, 0.45$^{+0.2}_{-0.2}$ for the \textit{g}-factor,
80$^{+40}_{-50}$ pc for the distance to the front face, and
70$^{+60}_{-50}$ pc for the thickness. The 90 $\%$ confidence ranges
for both the albedo and the \textit{g}-factor include the
theoretical values 0.4 for the albedo and 0.65 for the
\textit{g}-factor \citep{dra03}. It should be noted that the albedo
is well constrained with less than 10$\%$ error range for the 90$\%$
confidence. The distance and thickness values indicate that the dust
cloud is generally located in front of the OES, whose distance is
known to be $\sim$155 pc to the near-side. The wide ranges for the
distance and thickness estimated here may actually indicate the
large spatial dimension of the cloud along the line of sight, as we
will note in the Discussion section.


\section{Discussion}

As mentioned previously, the assumption of a single slab as a model
for the dust cloud in the OES is unrealistic, and the ranges of the
estimated distance and thickness were rather broad. Here, we would
like to improve the model by allowing local variations of the
distance and thickness of the cloud. However, we fix the value of
albedo to be 0.43, as we obtained with a single slab model. Being
the most sensitive one among the four parameters that determine the
model FUV intensity level, it was constrained very well. We also set
the value of the phase function asymmetry factor to be 0.45. The
range of \textit{g} was found to be rather broad, implying that the
model was not affected significantly by the change of \textit{g}.
Furthermore, Figure \ref{fig:contour}(d) and Figure
\ref{fig:contour}(f), in which the 90 $\%$ contours of the distance
and thickness were plotted against \textit{g}, do not show any
correlation with the value of \textit{g}. With these fixed albedo
and \textit{g}-factor values, we carried out simulations for the
sub-regions 1 to 9 individually to obtain the best fitting
parameters of the distance and thickness for each of the
sub-regions.

\begin{figure}
 \begin{center}
  \includegraphics[width=7.3cm]{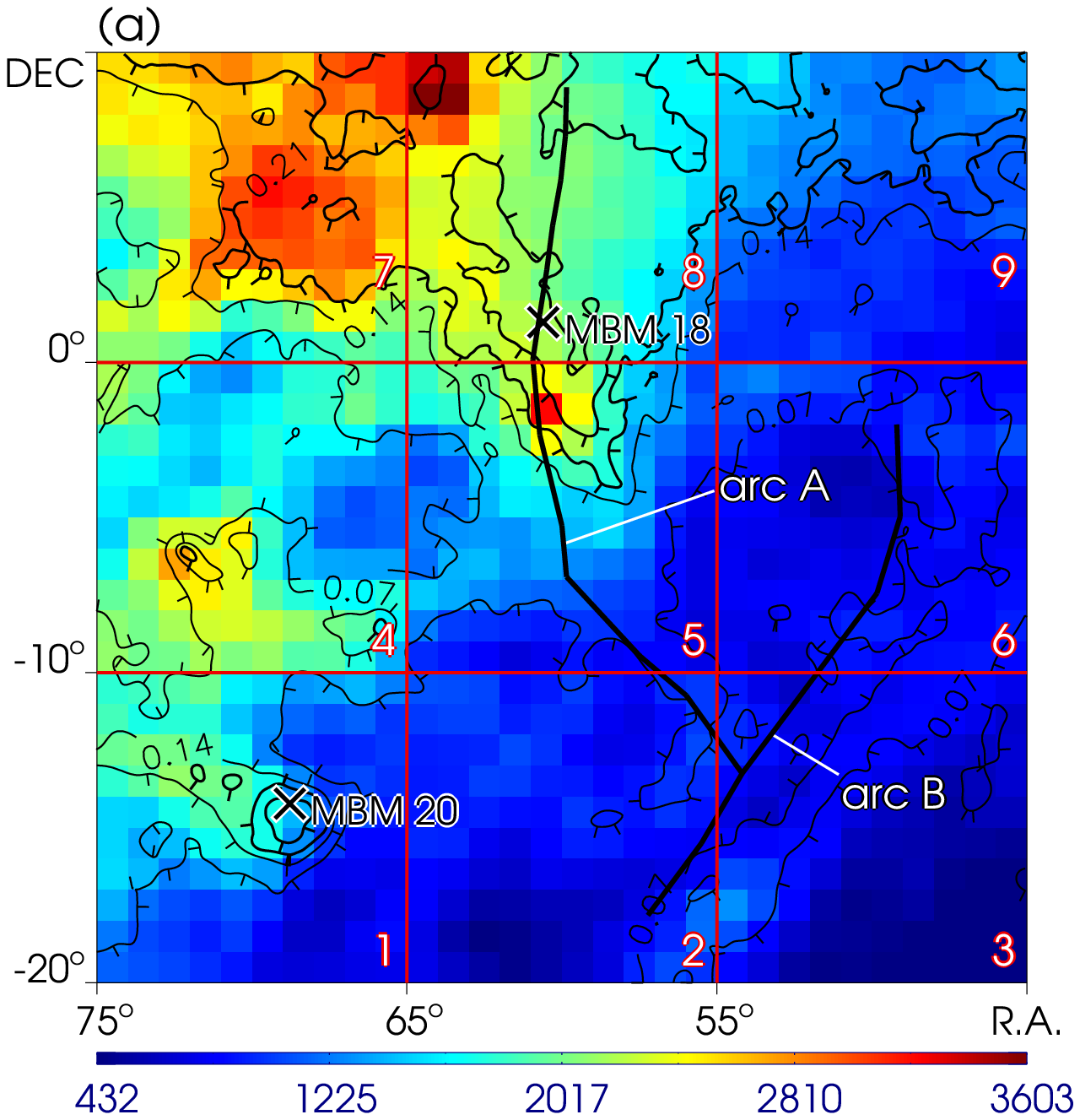}\\
  \includegraphics[width=7cm]{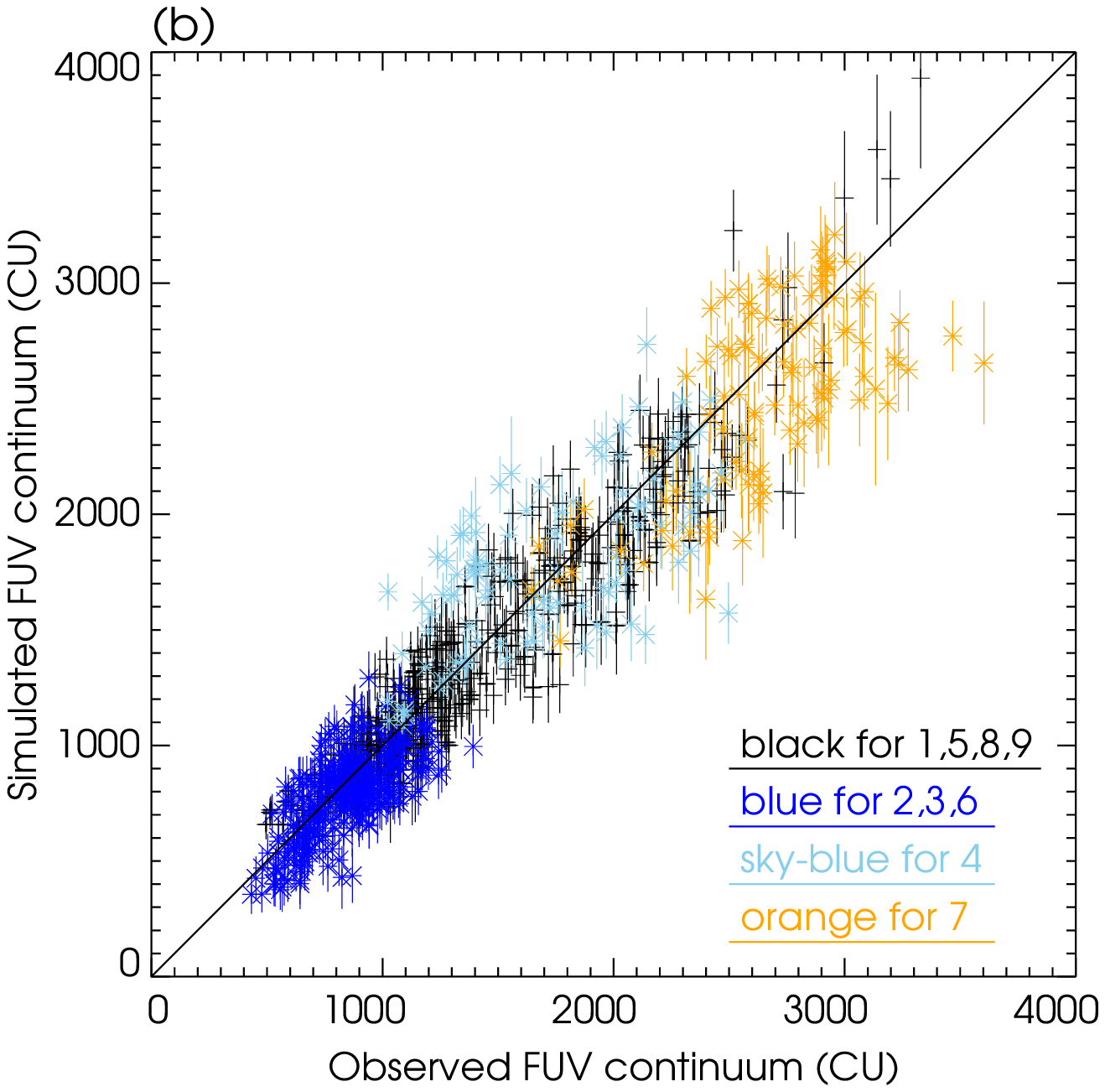}\\
 \end{center}
\caption{(a) A mosaic FUV continuum map obtained from dust
scattering simulations, with the distance and thickness of the
sub-regions varied individually, and (b) a pixel-to-pixel scatter
plot of the FUV intensities taken from the observed and simulated
maps in Figure \ref{fig:fims}(a) and Figure \ref{fig:simul2}(a),
respectively.\label{fig:simul2}}
\end{figure}

The best fitting values for the 9 sub-regions are shown in Table
\ref{tbl:simul2}, and the resulting mosaic map constructed with
these fitting parameters is depicted in Figure \ref{fig:simul2}(a),
together with the pixel-to-pixel scatter plot of the FUV intensities
in Figure \ref{fig:simul2}(b). As Figure \ref{fig:simul2}(b) shows,
the simulation result is much improved now when compared with the
previous result shown in Figure \ref{fig:simul}(b), with the new
reduced chi-square value of $\chi^2$ $\sim$ 3.19. The most
remarkable change is seen in the sub-region 7, of which the
intensity increased. Noting that the thickness increased
dramatically for this region, we believe the reason for the
intensity increase in sub-region 7 is that more stellar sources
reside in the cloud now and can contribute to the scattered
continuum. We also note that the contribution from the two-photon
effect, estimated as a difference of the FUV intensity between
Figure \ref{fig:fims}(a) and Figure \ref{fig:simul2}(a), has
decreased to $\sim$15-30$\%$, by $\sim$10$\%$ from the previous
result.

While the best fitting parameters shown in Table \ref{tbl:simul2}
may still not be exact, they seem to provide meaningful insights
into the morphology of the OES region. First, we note that the
sub-regions 1, 2, 3, and 6, corresponding to the blue regions, have
very thin ($\sim$10 pc) dust clouds located at $\sim$70-90 pc. As it
is believed that inner X-ray cavities of hot gas confront the
ambient dust clouds in the OES, the model result of thin dust layers
in these sub-regions may be the front boundary of the OES shell
encompassing the hot cavity. The clouds may have become thin as they
were blown out by the stellar winds or supernova explosions that
produced the hot cavity. Sub-region 1 includes the MBM 20 molecular
cloud \citep{mag85}, which \citet{bur93} placed at the interface
between the OES bubble and the Local bubble. While \citet{jo11} also
estimated its location to be in front of the OES bubble because the
cloud seemed to block the FUV emission from the OES, the present
model does not constrain its distance to be nearer than the clouds
associated with the OES in the sub-region, perhaps either because
the number of pixels relevant to MBM 20 is too small to affect the
simulation result or because there are no stellar sources that can
discriminate its distance from the OES clouds.

\begin{deluxetable}{cccccccccc} \tablecolumns{10} \tablewidth{0pc}
\tablecaption{Best fitting parameters for each sub-region with fixed
albedo 0.43 and g-factor 0.45 \label{tbl:simul2}} \tablehead{
  \colhead{sub-regions}
  & \colhead{1}
  & \colhead{2}
  & \colhead{3}
  & \colhead{4}
  & \colhead{5}
  & \colhead{6}
  & \colhead{7}
  & \colhead{8}
  & \colhead{9}}
\startdata
distance (pc) & 70 & 90 & 90 & 130 & 130 & 90 & 50 & 80 & 80\\
thickness (pc) & 10 & 10 & 10 & 10 & 70 & 10 & 130 & 70 & 50
\enddata
\end{deluxetable}

\begin{figure}
 \begin{center}
  \includegraphics[width=7.3cm]{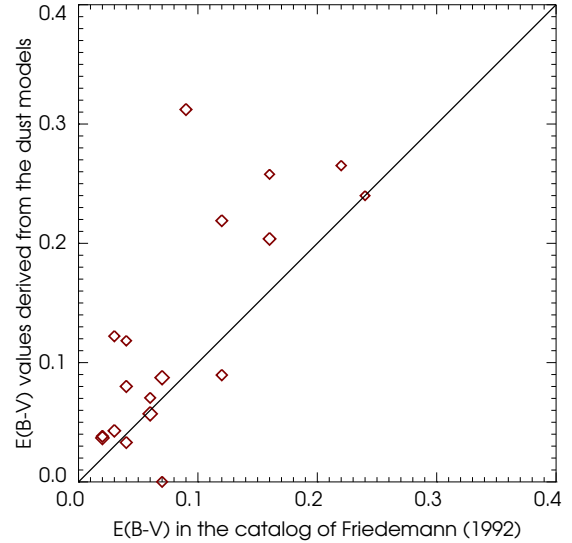}\\
 \end{center}
\caption{ \textit{E}(\textit{B-V}) values derived from the dust
models of the present study are compared with those for the 19 OES
stars listed in the catalog of \citet{fri92}.\label{fig:cmp_ebv}}
\end{figure}

Next, we note that the clouds of the sub-regions 7, 8, and 9 all
have their centers at a distance $\sim$110 pc, as the distance in
the table is actually the distance to the front face of the slab,
confirming it to be a single body. Hence, this part of OMCC seems to
be located at a distance $\sim$110 pc from the Sun, with its
thickness ranging from 130 pc at the core (sub-region 7) to 50 pc at
the boundary (sub-region 9). Sub-regions 4 and 5 are located between
the two distinct groups of dust clouds mentioned above and the
simulation results show quite different thicknesses for the two
sub-regions while their distances to the front face are similar. We
believe such a discrepancy arose because two different types of
structures, a thin shell surrounding the hot cavity and a thick dust
layer as part of the OMCC, coexist in these sub-regions. In fact,
when the simulation was carried out separately for the upper and
lower parts of these sub-regions, their thicknesses were
$\sim$100-140 pc for the upper parts and 10-30 pc for the lower
parts for both of the sub-regions, in accordance with the
environments of a thick dust region to the north and a thin shell to
the south. In reality, the structure of the dust clouds would be
more complex than we have assumed in the present study. However, we
believe that our results provide an overall structure of the dust
clouds toward the OES region.

With the new distance estimation for each of the sub-regions, we
would like to compare the \textit{E}(\textit{B-V}) values derived
from the distances of the dust layers with the observed
\textit{E}(\textit{B-V}) values for the OES stars listed in the
catalog of \citet{fri92}. While the derived \textit{E}(\textit{B-V})
values suffer from many error sources, as the present study modeled
very simple dust layers and the observed \textit{E}(\textit{B-V})
values are based on the stellar models that involve uncertainties,
we believe such a comparison should give insight into the validity
of our model. Figure \ref{fig:cmp_ebv} shows the result: the derived
\textit{E}(\textit{B-V}) values plotted against the observed
\textit{E}(\textit{B-V}) values. Although there are a couple of
outliers, a positive correlation can clearly be seen, with a
correlation factor of 0.75. However, the derived
\textit{E}(\textit{B-V}) is generally higher than the observed
\textit{E}(\textit{B-V}). While we may expect that real dust
distributions are not as simple as in the present study, such a
tendency may come from an over-estimation of the amount of dust
placed in front of the stars: in fact, we put the whole amount of
dust estimated from the SFD survey in the model layers, without
leaving any for the background.

Finally, it would be of great interest to know whether the
characteristics of the dust in the OES region are comparable to
those in previously studied regions, as the OES was suggested to
have been created by an energetic supernova explosion with an
additional energy source from the stellar winds produced by the
stars in the I Ori association \citep{rey79}, and the origin of the
OES might be reflected in the properties of the dust. It has been
shown that dust grains characterized by relative flat FUV extinction
curves stand out with relatively high FUV albedo values, in the
cases such as IC 435 \citep{cal95} and ScoOB2 \citep{gor94}, while
dust with steeper far-UV extinction curves, suggestive of a larger
component of small grains, exhibits lower albedo values
\citep{wit97}. We fit each of the 9 sub-regions by treating the
albedo as a free parameter to see if there is a difference between
the thick dust region associated with the Orion Molecular Cloud
Complex in the north and the cavity region where hot gas prevails.
However, we could not find any systematic changes in the albedo
among the sub-regions, with its values varying from 0.42 to 0.47. We
also checked the R$_V$ values to see if there is a systematic
variation among the sub-regions using the data from \citet{lar05}
and the A$_V$ values provided by \citet{nec80}, as mentioned in
Section 3, but could not find a meaningful change between the
northern molecular cloud region and the southern cavity region,
although statistics may not be high enough as there are only two
stars located in the cavity region. Hence, the dust characteristics
obtained in the present study seem to be similar to those of the
diffuse ISM, with no exceptional signatures in the albedo and the
R$_V$ values. We believe the reason is probably that most of the
dust seen in the present study is located outside the OES. In fact,
the structure of the dust layer estimated in the present study, with
its thickness of 10 pc at a distance of 90 pc in the cavity regions,
is in good agreement with this view of it as a thin layer between
the OES and the Local Bubble.


\section{Conclusions}

We have performed Monte Carlo simulations for the dust scattering of
FUV emissions in the OES region and compared the results with the
diffuse emission map made from the recent FUV imaging spectrograph
mission, FIMS. We were able to obtain the optical parameters of
interstellar dust grains for the OES region: the average albedo is
0.43$^{+0.02}_{-0.04}$ and the phase function asymmetry factor is
0.45$^{+0.2}_{-0.2}$, in agreement with previous observational and
theoretical estimations. Furthermore, the simulation results
indicate that the dust clouds are located in front of the OES, with
its distance of $\sim$110 pc and thickness of $\sim$50-130 pc, while
the hot X-ray cavity is bounded by a thin ($\sim$10 pc) dust shell
toward the Sun.

\acknowledgments

FIMS/SPEAR is a joint project of KAIST and KASI (Korea) and UC
Berkeley (USA), funded by the Korea MOST and NASA grant NAG5-5355.
This research was supported by Basic Science Research Program
(2010-0023909) and National Space Laboratory Program (2008-2003226)
through the National Research Foundation of Korea (NRF) funded by
the Ministry of Education, Science and Technology.

\clearpage

\end{document}